\documentclass[iop]{emulateapj}		
\usepackage{epsfig}
\usepackage{epstopdf}			
\usepackage{graphicx,color}		
\usepackage{amssymb}			
\usepackage{color}			
\usepackage{url}			
\usepackage{amsmath}			
\usepackage{rotating}			
\usepackage{float}			
\usepackage{textcomp}			
\usepackage{psfig}
\usepackage{dcolumn}
\usepackage{times}
\usepackage{tabularx}
\usepackage[colorlinks=true,citecolor=blue]{hyperref}
\usepackage[english]{babel}

\shorttitle{Propagating Disturbances in the Solar Corona and  Spicular Connection}
\shortauthors{T. Samanta et al.}

\begin{document}

\title{Propagating Disturbances in the Solar Corona and  Spicular Connection}

\author{Tanmoy Samanta$^{1}$,
Vaibhav Pant$^{1}$,
Dipankar Banerjee$^{1, 2}$}

\affil{$^{1}$Indian Institute of Astrophysics, Koramangala, Bangalore 560034, India.\\ 
{$^{2}$ Center of Excellence in Space Sciences, IISER Kolkata, India }\\e-mail: {\color{blue}{tsamanta@iiap.res.in}}\\}

\begin{abstract}
Spicules are small hairy like structures seen at the solar limb mainly at chromospheric and transition region lines. 
They generally live for 3-10 minutes. We observe these spicules in a south polar region of the Sun with a coordinated observations using 
the Interface Region Imaging Spectrograph (IRIS) and the Atmospheric Imaging
Assembly (AIA) instruments on board the Solar Dynamics Observatory. 
Propagating disturbances (PDs) are observed everywhere in the polar off-limb regions of the Sun at coronal heights. 
From this simultaneous observations we show that the spicules and the PDs may be originated by a common process. 
From space-time maps we find that  the start of the trajectory of PDs is almost co-temporal with 
the time of the rise of the spicular envelope as seen by IRIS slit-jaw images at 2796~\r{A} and  1400~\r{A}. 
During the return of spicular material, brightenings are seen in AIA 171~\r{A} and 193~\r{A} images.
The quasi-periodic nature of the spicular activity as revealed by the IRIS spectral image sequences and 
its relation to coronal PDs as recorded by the coronal AIA channels suggest that they have a common origin. 
We propose that reconnection like processes generate the spicules and waves simultaneously. 
The waves escape while the cool spicular material falls back.
 
\end{abstract}
\keywords{Sun: oscillations --- Sun: corona --- Sun: transition region --- Sun: UV radiation --- Sun: magnetic topology}


\section{Introduction}
White light and Extreme Ultra Violet (EUV) emission from the polar region of the solar corona shows distinct bright ray like structures 
known as polar plumes  \citep{1950BAN....11..150V,1958PASJ...10...49S,1965PASJ...17....1S,1997SoPh..175..393D,1998ApJ...501L.217D}.
Quasi-periodic propagating intensity perturbations are frequently observed in the polar plumes and inter-plume regions with period ranging from 5--30 minutes. 
The measured speed of these propagating disturbances (PDs) is around 150 km s$^{-1}$.
These outward  PDs are generally interpreted as slow magneto-acoustic waves propagating through the plumes and inter-plumes region  because their speeds are similar to the sound speed
\citep{1998ApJ...501L.217D,1999ApJ...514..441O,2000ApJ...533.1071O,2000SoPh..196...63B,2007A&A...463..713O,2009A&A...499L..29B,2011SSRv..158..267B,2011A&A...528L...4K,2012A&A...546A..93G,2014A&A...568A..96G,2014ApJ...793..117S,2015arXiv150504475B}. 

EUV observations from Solar TErrestrial RElations Observatory
(STEREO), \citet{2010A&A...510L...2M} observed high speed 
jets of plasma traveling along the plume structures with a mean velocity around 135 km s$^{-1}$. Those jets were quasi-periodic with the periodicity of 5--25 minutes. \citet{2011ApJ...736..130T} also showed that high-speed repetitive outflows originate near magnetic network elements both in the quiet-Sun (QS) and coronal holes (CHs) propagate along plume structures with an average speed $\sim$~120 km s$^{-1}$.
\citet{2010A&A...510L...2M} suggested that the PDs are nothing but these jets originating from the upper chromosphere propagate to higher corona and contribute to the solar wind. They further conjectured that these jets will supply adequate energy to the fast solar wind. \citet{2014ApJ...793...86P} found radially moving radiance variations both in
the plume and inter-plume regions. Comparing the apparent outflow speeds at different temperature passbands they concluded that the observed radiance variations represent the material outflows.

The quasi-periodic intensity enhancement, the Doppler shift and the line width enhancements in coronal structures are also recently interpreted as due to high-speed upflows from spectroscopic
observations in magnetic regions of the solar atmosphere \citep{2009ApJ...701L...1D,2010ApJ...722.1013D,2011Sci...331...55D,2011ApJ...727L..37T,2011ApJ...738...18T,2011ApJ...732...84M,2012ApJ...759..144T}. \citet{2015ApJ...807...71P} have studied an on-disk plume as seen in AIA and their connection to transition region jets seen in IRIS. 
They found that small scale chromospheric jet-like features are linked with the generation of
PDs within the plume. \citet{2011Sci...331...55D} found that spicules seen in AIA 304~\r{A} channels propagate upward and fall back following  parabolic paths whereas at the same time they also found that 
enhanced emissions seen in 171~\r{A} propagate upward into the corona 
with speeds of $\sim$~100 km s$^{-1}$.
If similar behavior can been seen in the off-limb corona is yet to be explored and is the subject of the present analysis.

Recently \citet{2015ApJ...809L..17J} claimed that the spicular activities in the solar transition region, as seen in the AIA 304~\r{A} passband, 
are responsible for the generation of the PDs in the polar regions of the corona as observed in the AIA 171~\r{A} images. 
However, the exact connection between spicule and PDs and on the nature of the PDs still remains a mystery. 
Furthermore for the solar wind studies it is important to have a better understanding on the nature of these PDs and their origin. 
To find out the spicular connection to the coronal PDs, we use simultaneous IRIS data along with coronal data from AIA.


\section{Data analysis and Results}
\subsection{Observation and Data Reduction}
\begin{figure*}
\centering
\includegraphics[angle=90,clip,width=17.5cm]{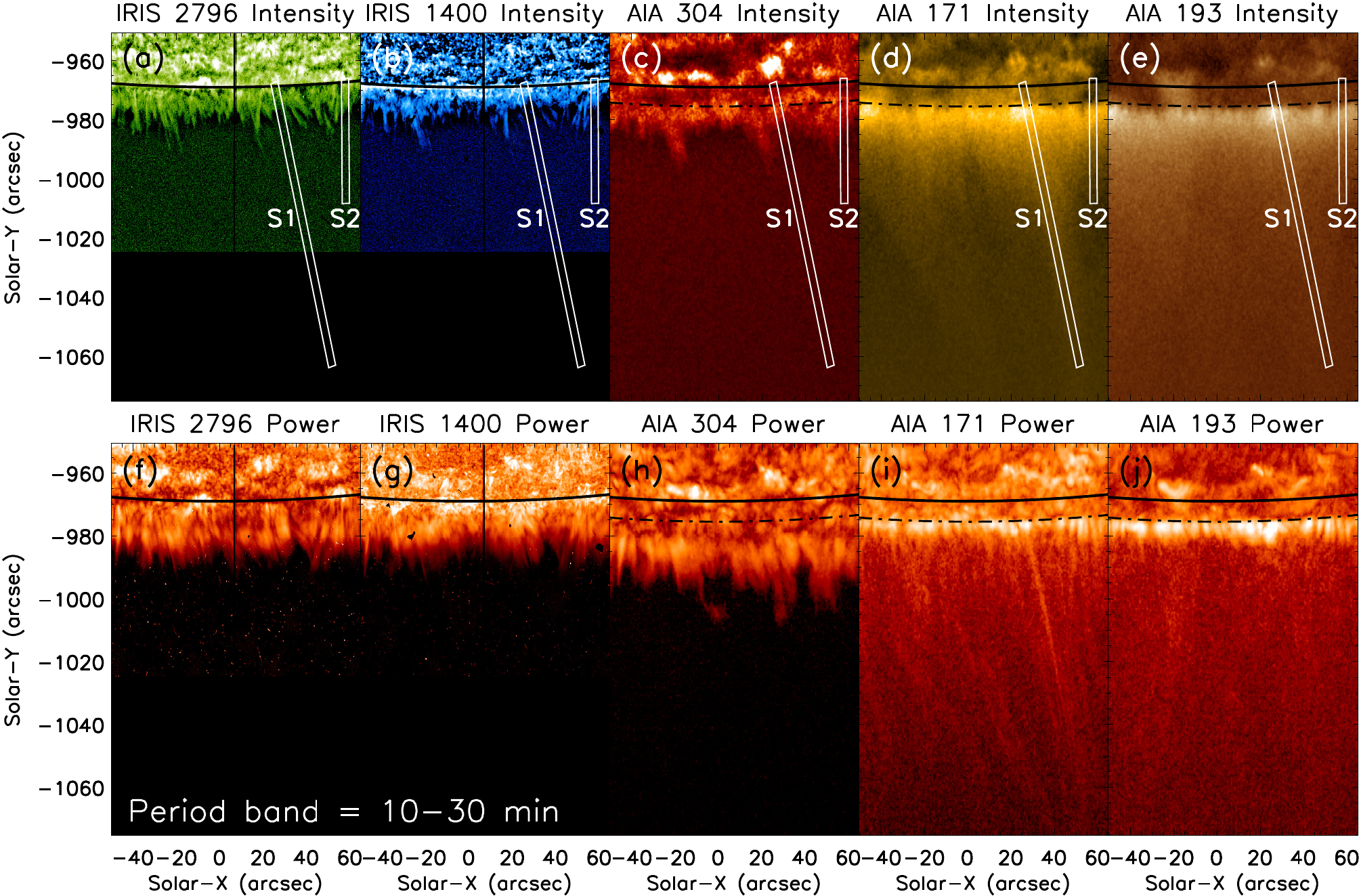}
\caption{(a)-(e) display a portion of the south polar region of the Sun as seen in different filtergram images taken from IRIS and SDO/AIA instruments
on 21 February 2014 as marked (an animation is available online: movie1). 
The vertical black line on each IRIS SJIs represents the position of the IRIS slit. The solid black curve line on each image shows the location of the solar limb. 
The dot-dashed black line on each AIA image display the location of the solar limb as identified from the AIA 171~\r{A} image. 
The field of view (FOV) of IRIS is smaller than 
selected window of AIA images which makes it appears dark in SJIs where IRIS FOV does not overlap with AIA. 
Two boxes (S1 and S2) shows the location of selected slits used to construct space-time (X-T) plot (see Figure~\ref{xt_s1} and \ref{xt_s2}). 
(f) -- (j) show distribution of power (powermaps) of corresponding top panel channels. The powermaps are constructed by taking average power of 10--30 minutes period band.} 
\label{aia_iris} 
\end{figure*}
The data was obtained from a coordinated observations using the Interface Region Imaging Spectrograph (IRIS) satellite \citep{2014SoPh..289.2733D} and the Atmospheric Imaging
Assembly (AIA) instruments \citep{2012SoPh..275...17L} on board the Solar Dynamics Observatory (SDO). Observation was performed from 11:25~UT to 12:55~UT on 21 February 2014. 
Slit-jaw images (SJI) from IRIS  filtergrams centered at 2796~\r{A} and 1400~\r{A}, dominated by  \ion{Mg}{2}  and \ion{Si}{4} emission line, respectively, were analyzed.
AIA filtergram images centered at 304~\r{A}, 171~\r{A}, and 193~\r{A}, dominated 
by \ion{He}{2}, \ion{Fe}{9} and \ion{Fe}{12} emission respectively, were also selected for the analysis. 
IRIS 2796~\r{A} passband is sensitive to emission from plasma at temperatures $\sim$10,000 to 15,000 K and IRIS 1400~\r{A} is sensitive to temperatures $\sim$ 60,000 to 80,000 K.
The AIA 304~\r{A}, AIA 171~\r{A} and AIA 193~\r{A} filters response function peaks at 0.05~MK, 0.8~MK and 1.25~MK respectively.

IRIS observed south pole of the Sun in sit-and-stare mode.
We used IRIS Level~2 processed  data which is corrected for
dark current, flat field and geometrical corrections etc. 
All the IRIS SJI were taken with 8~sec exposure time and a cadence of 19~sec. 
The pixel size of SJI's and AIA are $0.166''$ and $0.6''$ respectively. The AIA images were taken with a 12 sec cadence.
Pixel size of AIA is interpolated to the match with IRIS for easier comparison.
All the AIA channels were co-aligned and de-rotated to compensate for solar rotation.
The IRIS SJIs and AIA images were co-aligned using IRIS 1400~\r{A} and AIA 1600~\r{A} images (as described in \citet{2015ApJ...806..170S}).

Figures~\ref{aia_iris}~(a)--(e) show the south polar region of the Sun as seen from IRIS and AIA channels. 
We should point out here that  coronal hole was not seen clearly on the south pole during the
observation, hence the the region can be characterized as quiet Sun. Furthermore, distinct extended plume structures were not so clearly visible from original images. Usually when there is a underlying deep coronal hole present, plume structures are clearly visible.

\begin{figure*}
\centering
\includegraphics[angle=00,clip,width=17.5cm]{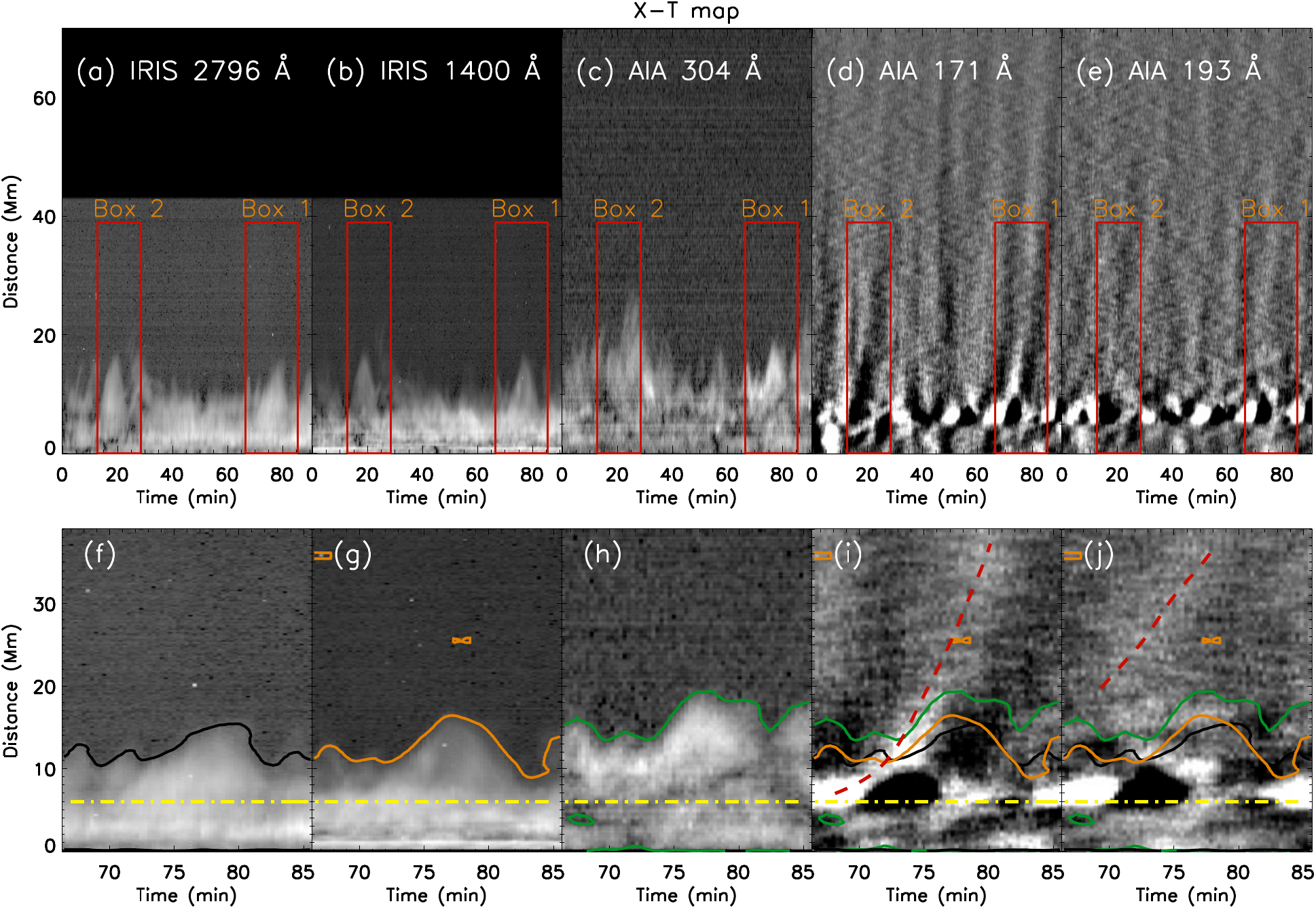}
\caption{X-T plots corresponding to S1 slit and different passbands as marked.
A zoomed view of the region inside the red rectangular Box~1 (marked in the upper panels) is shown in (f) -- (j). Animations corresponding to Box~1 (movie2$_{-}$box1)~\&~2 (movie3$_{-}$box2) is available online. 
Black, orange and green contours show the envelopes of the spicular temporal evolution as seen in the IRIS 2796~\r{A} SJI, IRIS 1400~\r{A} SJI and AIA 304~\r{A} channels. 
These contours are overplotted in the AIA 171~\r{A} and AIA 193~\r{A} X-T map for comparison. 
Dot-dashed horizontal line in yellow represents the limb of Sun as seen in AIA 171~\r{A} and 193~\r{A}. 
Dashed red curves track the PDs observed in AIA 171~\r{A} and 193~\r{A}.}
\label{xt_s1} 
\end{figure*}
\begin{figure*}
\centering
\includegraphics[angle=0,clip,width=17.5cm]{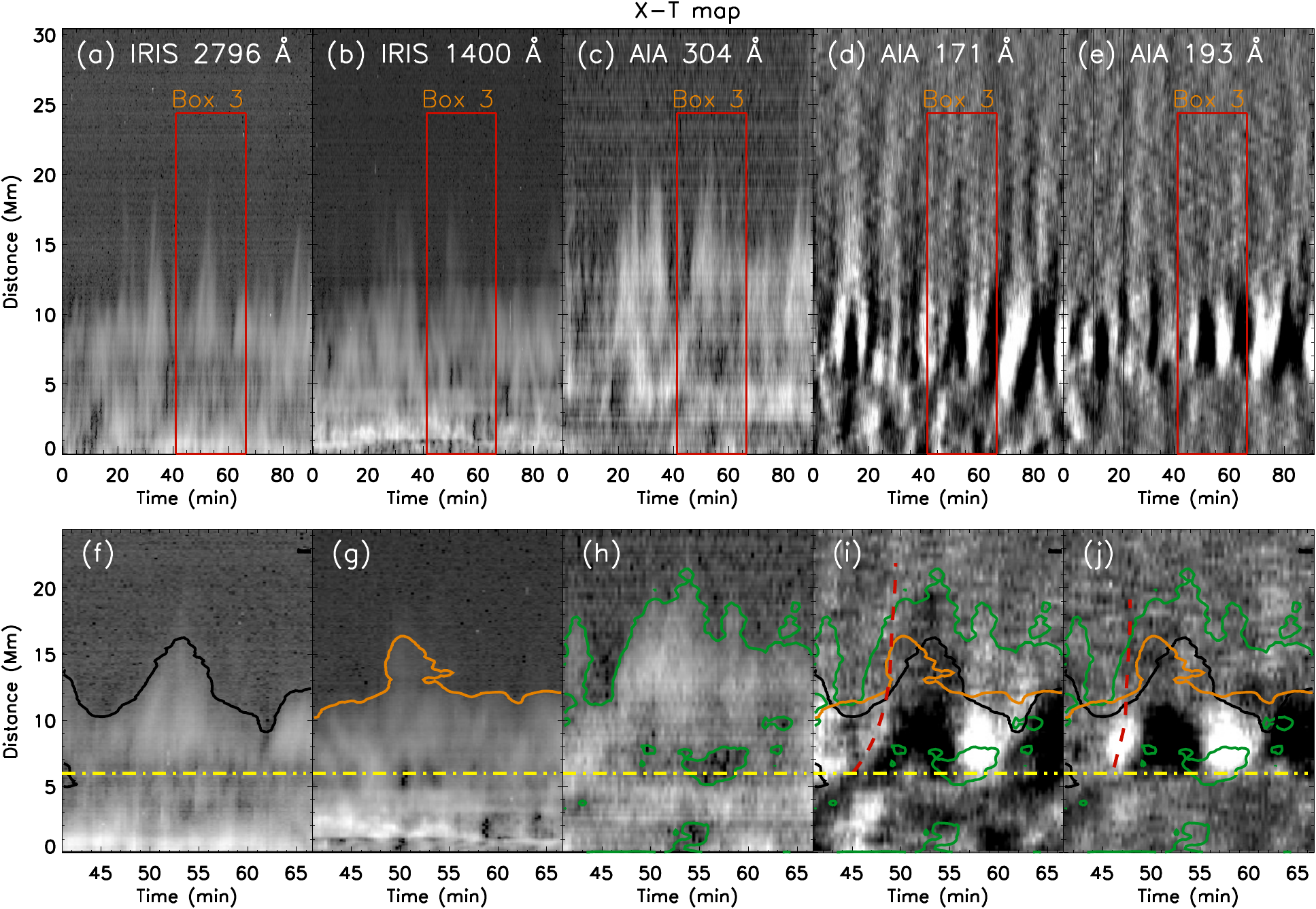}
\caption{X-T plots corresponding to the S2 slit  and different passbands as marked. 
All panels are similar to those in Figure~\ref{xt_s1}. Animation corresponding to Box~3 (movie4$_{-}$box3) is available online.} 
\label{xt_s2} 
\end{figure*}

\subsection{Power Maps}
PDs in the corona often repeat with a time scale of 10--30 minutes.
To find the global behavior of PDs, we constructed power maps in the period band of 10--30 minutes.
We performed wavelet analysis in time series at each
pixel location to estimate the distribution of power. A background trend
is removed by subtracting a 30 minutes running average of the original time series to remove periods longer than 30 minutes.
Figure~\ref{aia_iris}~(f)--(j) show the distribution of power (referred to as powermaps) within the 10--30 minutes period band of IRIS 2796~\r{A}, IRIS 1400~\r{A}, AIA 304~\r{A}, AIA 171~\r{A} and
AIA 193~\r{A} channels. The powermaps of IRIS 2796~\r{A}, IRIS 1400~\r{A} and AIA 304~\r{A} show the power in this period band is limited to lower heights (within $\sim$20$^{''}$ from the solar limb).   
These channels are sensitive to the chromospheric and transition region temperature which is dominated by spicular activity and probably due to this the power is limited to lower heights. 
At some places we note that the strong power is seen up to higher heights.  In AIA 171~\r{A} and AIA 193~\r{A} powermaps, we observed two kinds of features, one reaching to
larger distances (see the location of S1 slit) and other which have comparatively higher power but confined to shorter distances (see the location of S2 slit).
In the next subsection we will study in details the behaviour of PDs in these regions (within S1 and S2).

\subsection{Space-Time Plot}
In this subsection, we focus on the plume like structure marked by the S1 slit in Figure~\ref{aia_iris}
for a detailed time evolution of PDs. We constructed X-T map to study the temporal evolution of the region inside the S1 slit. 
The signal in the AIA channels is low for off-disk features, hence to improve the signal to noise, we have used a thick slit of width of 3.32$^{''}$. 
The average intensity along the width of the slit was used to
construct the X-T maps of different channels. The X-T map of AIA 171~\r{A} and 193~\r{A} were processed by removing a smoothed background trend along time axis 
to enhance the visibility of the alternating ridges. 
The results are shown the Figure~\ref{xt_s1}. The X-T maps of the IRIS 2796~\r{A}, IRIS 1400~\r{A} and AIA 304~\r{A} which are sensitive to 
chromospheric and transition region temperatures show evolution of several spicules. On the other hand the
dark and bright ridges extended over longer distances are seen in the AIA 171~\r{A} and 193~\r{A} channels. The zoomed view of a portion of the X-T map is shown in the bottom
panels of the same Figure. The temporal evolution of spicules are seen in IRIS 2796~\r{A}, IRIS 1400~\r{A} and AIA 304~\r{A}  appears to have sub-structures rising and falling in all passbands. 
It appears that they roughly follow parabolic path as seen in many spicules. Contours are over-plotted by choosing some intensity threshold on the IRIS 2796~\r{A}, IRIS 1400~\r{A} and AIA 304~\r{A} channels 
which are shown in black, orange and green contours respectively. It covers the envelope of the spicular temporal evolution. These three contours are also overplotted in the
AIA 171~\r{A} and 193~\r{A} X-T maps. The trajectory of the PDs are marked with red dashed curve as shown in zoomed AIA 171~\r{A} and 193~\r{A} panels of Figures~\ref{xt_s1} and ~\ref{xt_s2}.

Similar procedure was adopted for slit S2 analysis. The slit width of the slit S2 is same as S1. 
The X-T maps for different channels are shown in Figure~\ref{xt_s2}.
In the lower panels of Figures~\ref{xt_s1} and ~\ref{xt_s2}, we note that the start of trajectory of PDs is almost co-temporal with the time of the rise of the spicular envelope. 
We also note that the falling of the spicular envelope is followed by brightenings and generation of another PD in AIA 171~\r{A} and 193~\r{A}.

It is worth noting that PDs in Figure~\ref{xt_s1} (following a plume like structure) are propagating to higher heights (65 Mm) as compared to PDs in Figure~\ref{xt_s2} 
which seems to propagate roughly about 30 Mm. Thus, power was confined to lower heights at slit S2 location.


\begin{figure*}
\centering
\includegraphics[clip,width=11cm]{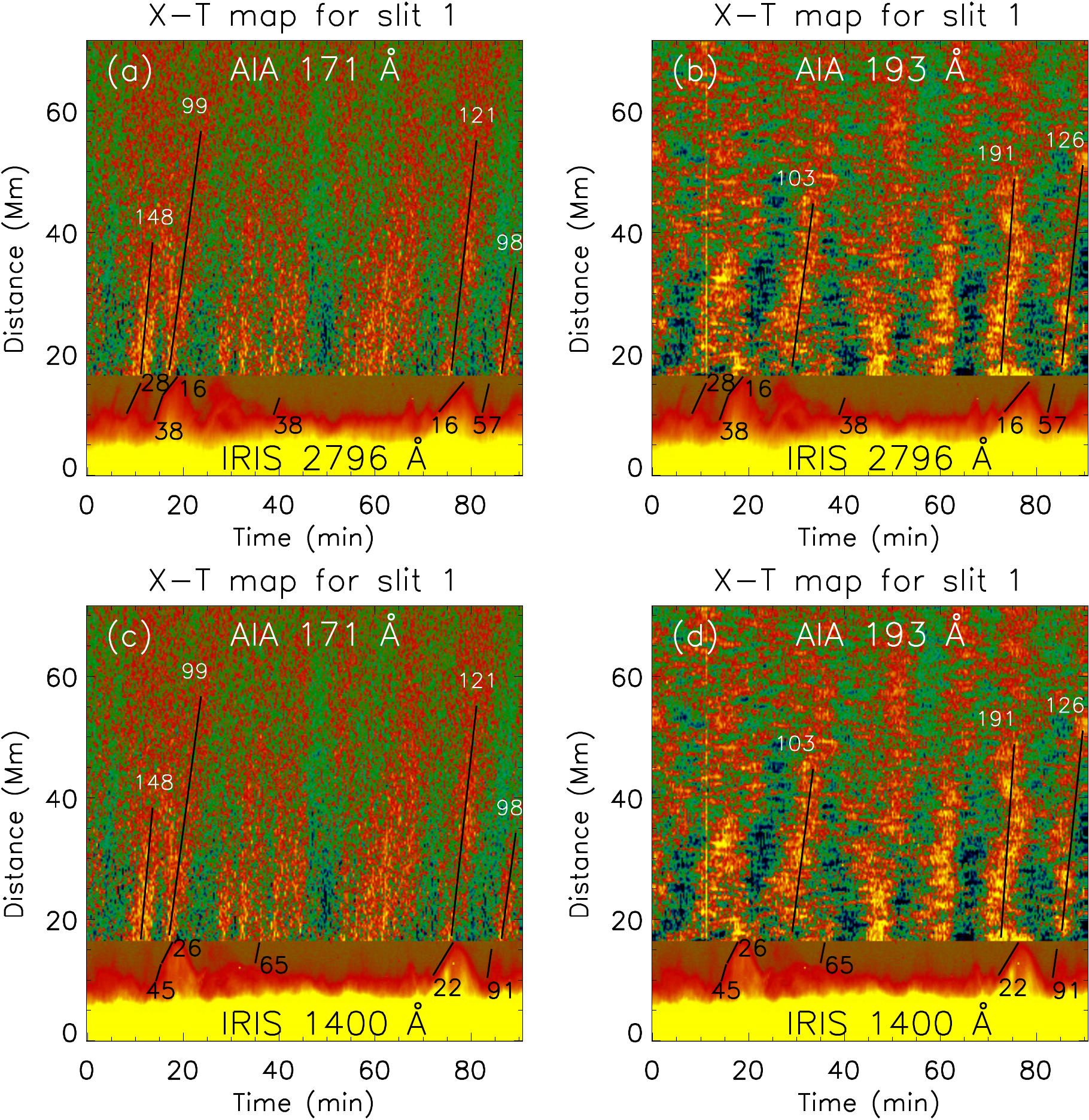}
\caption{X-T corresponding to slit location S1 (along a plume like structure) showing connection between spicular activities as seen by IRIS channels (bottom) and PDs as seen by coronal AIA channels (top). 
The slanted black lines are used to measure the speeds. Measured speeds in km~s$^{-1}$ are printed.}
\label{vel2} 
\end{figure*}

\subsection{Measurements of propagating speed of the PDs}

To follow the PDs and its connection with rapidly evolving spicular activities we produce composite X-T plot (Figure~\ref{vel2}). The IRIS channels corresponding to the bottom part of the maps (as marked) allows to identify the spicules and its temporal evolution whereas the AIA channels corresponding to the upper part of the maps allow us to identify the PDs quite clearly. In order to see the alternating ridges clearly we enhance the contrast by subtracting the smooth background in X-T maps. 
This procedure is followed for AIA channels (i.e, upper part of the maps in Figure~\ref{vel2}). For IRIS channel (lower part of the maps) no background subtraction is performed because no enhancement in lower part of X-T maps is needed. 
It further demonstrates that when a spicule is observed by the chromospheric channel of IRIS a PD seems to be generated and/or amplified in the AIA coronal channels. 
This also confirms that the PDs seems to be quasi-periodic with a periodicity which is governed by the repetition time scale of occurrence of the spicules. 
From this figure we also fit the ridges manually using eye estimation and found the range of velocities from 98 - 148 km s$^{-1}$ in AIA 171~\r{A} and 103 -- 191 km s$^{-1}$ in AIA 193~\r{A}. 
Since there is a large uncertainty in the measured speed thus we can not confirm if we observe different speeds in the two different channels. 
We find the speeds of spicules from 22 -- 91 km s$^{-1}$ corresponding to IRIS 1400~\r{A} images and 16 -- 57 km s$^{-1}$ for IRIS 2796 ~\r{A} images. 
Since the speeds are higher than sound speed at chromospheric levels these spicules could be generated  by reconnection at lower heights. 

\section{Conclusion}

PDs are ubiquitous in the solar atmosphere and recently it has been suggested by \citet{2015ApJ...809L..17J} that spicules can trigger coronal PDs. 
From their study using the AIA data alone they provided a possible link of spicular activities with the generation of PDs. They used the AIA 304~\r{A} channel to identify the spicules. 
They did not comment on the nature of the PDs as well. We extended that work using the higher cadence and higher resolution IRIS spectral images to 
study the dynamical behaviour of the spicular activity and while combining AIA channels we are able to confirm the link with a better confidence. 
We feel a combination of chromospheric and transition region channels of  IRIS and transition region and coronal channels of AIA is better suited for this kind of study.
We also point out a possible connection between the small scale, short lived cool spicular structures with the very large scale long lived hot plume structures. 
We find that the start of trajectory of PDs is almost co-temporal with 
the time of the rise of the spicular envelope and the spicular material fall is co-temporal with brightenings followed by the generation of another PD in AIA 171~\r{A} and 193~\r{A}.
Co-temporal generation of spicules and PDs suggest that they might have a common origin. 
We should point out here that there could be a projection effect and we can not rule out the possibility that the spicules and 
PDs may not be in the same position but statistically we find co-temporal and co-spatial match in  most of the cases studied.
Figure~\ref{vel2} illustrates the quasi-periodic 
nature of the spicular activity as revealed by the IRIS spectral image sequences and its relation to coronal PDs as recorded by the coronal AIA channels. 
We propose that reconnection like processes generate  the spicules and waves simultaneously. The waves escape while the cool spicular material falls back. 
The PDs as seen have a mixed signature of waves and flows at lower heights while at the extended coronal heights the wave signature is dominant. 
There is also a signature of acceleration of these PDs as represented by the curved red dashed lines in 
the lower panels of Figures~\ref{xt_s1} \& ~\ref{xt_s2}. 
The nature of PDs in our case is not always confirmatory but, 
presence of alternating bright and dark ridges and reaching of accelerating PDs to higher 
heights in slit position S1 (see Figures~\ref{xt_s1},~\ref{vel2}) suggest more wave like nature while manual fitting of 
PDs to derive speeds is not accurate enough to confirm wave like behaviour.
We should also point out that we did not have any coronal holes in the polar regions and thus the presence of plume like structures were also not very clear at all locations. 
At slit position S2 we find that PDs are not going to higher heights. Thus, power is confined to lower heights close to the limb. 
The enhancement in power is due to repeated generation of cool spicular material seen as dark features in hot AIA channels. Due to almost straight ridges, 
poor signal to noise, we could not measure the speeds of these PDs convincingly and thus can not comment on their nature. 
It is also possible that at the time of the generation of spicules (due to heating together 
with reconnection and/or driven by p-modes) the hot material can escape following the open magnetic field lines and thus cause intensity enhancements 
in the AIA 171~\r{A} and 193~\r{A} channels while spicular material falls back because it is denser and cooler than escaping hot material.
With the existing spectral imaging data the connection of spicules and PDs are subject to line of sight uncertainties. 
For further confirmation and better understanding on the nature of the PDs, we need simultaneous chromospheric and coronal long spectral time sequences. \\


\acknowledgments
We thank the IRIS team for proving the data in the public domain. We also thank the referee for valuable comments.
IRIS is a NASA small explorer mission developed and operated by 
LMSAL with mission operations executed at NASA Ames Research center 
and major contributions to downlink communications funded by the Norwegian Space Center (NSC, Norway) through an ESA PRODEX contract.


\end{document}